# Information criminality – a phenomenon met within the informatics field

Filote C., Nemțoi G.

**Abstract**—The phenomenon described as "information criminality" has taken significant proportions in the last decade, fact that carried out towards an international legislative frame, by implementing judicial forms, which might stop its occurrences. As matter of fact, the information criminality represents an information technology aiming towards fraud and prejudicing the users of informational data, by various means to infringement of the law. In this way, some international organizations have dealt with performing a legislative framework, able to punish the phenomenon of information criminality and implicitly to protect the users of computers. The transnational expansions, extremely fast as concerns the computer networks, and extending the access to these networks, by means of mobile telephony, have brought the increasing of these systems' vulnerability and the creating of opportunities of breaking the law. Considering these aspects, the world legislation is continuously changing, due to a more and more accelerated development of the information technology.

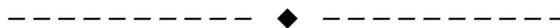

## 1 INTRODUCTION

THE progress of information technology field, taken into consideration at the world level, has created a framework of high potential for producing some antisocial actions called "criminal", both for international criminology and for informatics. The current computing systems are offering an adequate frame in order to commit the crimes concerning the *information criminality*, where most of social life services, such as controlling the flight traffic, trains traffic, coordination of medical services or international security, have been developed upon basis of informational programs.

The transnational expansions, extremely fast as concerns the computer networks, and extending the access to these networks, by means of mobile telephony, have brought the increasing of these systems' vulnerability and the creating of opportunities of breaking the law.

Considering these aspects, the world legislation is continuously changing, due to a more and more accelerated development of the information technology, where the international cooperation is confronting with an increasing aspect of the last decade, related to fighting against the information criminality and the eradication of this phenomenon.

The main issues that were discussed within the international meetings upon fighting against the information criminality were the following:

- the lack of a global consensus as concerns the definition of "information criminality",
- the lack of a global consensus regarding the occurrence of these actions,
- the lack of expertizations coming from authorized representatives of some institutions, having attributions of auditing in the field,
- the inexistence of some regulations, proper to the access and the investment in information systems, including the lack of the rules by which the computerized data basis might be confiscated,
- the lack of legislative harmonization as regards the means of the methodology of investigations in the field,
- the transnational character of this type of infringement of law,
- the existence of a reduced number of treaties or international agreements between states, which enacts the extradition and mutual assistance in the field.

One of the first international organizations that have discussed the topic approached of "information criminality" is represented by the Organization for Cooperation and Development (OECD).

This international body has accomplished a study concerning the harmonization of the legislation in the field since 1983, when by means of a report issued; the organization proposed to the EU member states a list of activities, which should support some sanctions – meaning the fraud counterfeiting performed by the help of a computer, such as: the alteration of computing programs, data or copyright, interception of communications or of other computing functions, the access and the unauthorized using of a computer, etc.

Towards the filling of a report issued by OECD and by the help of a commission of experts in criminality field,

---

• *Filote C. is with the Computers and Automation Department, Faculty of Electrical Engineering, Stefan cel Mare University of Suceava, 13 University Str., Suceava, Romania.*

• *Nemtoi G. is with the Law and Judicial Sciences Department, Faculty of Economics and Public Administration, Stefan cel Mare University of Suceava, 13 University Str., Suceava, Romania.*

The Council of Europe has issued a Recommendation R (89)9, concerning the criminality specific to unauthorized computer using, fact that indicates to national legislative structures the leading principles in order to define some crimes; in this way, the recommendation will signify an action guide for the EU member states. The Council of Europe issued on 23rd of November 2001 a Convention regarding the informational criminality.

The normative deed published in 2003 and issued by the Council of Europe has the aim of completing the normative deeds previously issued, in the view of increasing the efficiency of criminal investigations and procedures, having the main topic upon crimes related to information systems and data, as well as allowing the collection of electronic evidences.

The Resolution no.1, adopted by the European ministers of justice at the XXIst Conference of Prague, recommended the supporting of activities related to fighting against information criminality, in other words, assuring of collaboration between national criminal laws and using of some efficient means to the investigation of information crimes; the same topic refers to Resolution no.3, adopted at the XXXIIIrd Conference of European ministers, London 2000, which encourages the participating parties towards negotiations, and to continue the efforts in order to find solutions.

These solutions have to allow a highest number of states of becoming part of the convention, by recognizing the necessity of being provided with a fast and efficient mechanism of international cooperation, able to take into account the specific exigencies of the fight carried out against information criminality.

Simultaneously, another international organization, meaning Organization of United Nations, reached a warning sign towards the crimes that might occur, by means of computer networks. This warning sign was emphasized by publishing many documents, amongst which: "Proposals as concerns the focusing of international actions over the fight against any form of criminal activity" (1985), The Resolution introduced by the representative of Canada, as regards the fight against criminality by using computers (1990), "The Declaration of United States concerning the basic principles of the law applicable to victims of ultra vires actions or crime" (1990), The Report "Challenging without frontiers: cybercrime – international efforts for fighting against organized or transnational crimes" (2000).

## 2 THE PHENOMENON OF INFORMATION CRIMINALITY WITHIN NATIONAL AND INTERNATIONAL LEGISLATION

The European Council has initiated many approaches towards the establishment of activities on cybernetic area, precisely for stopping the phenomenon of information criminality. Taking into account a chronological order, one might notice:

- recommendation R(85)10, including rules of applying The European Convention on Mutual Assistance in Criminal Matters, as regards the rogatory committees on intercepting the telecommunications,
- recommendation R(88)2 regarding the piracy within the existence of copyrights and connected rights,
- recommendation R(87)15 specific to regulation of using personal data on police work,
- recommendation R(95)4 as concerns the protection of data within telecommunication services,
- recommendation R(95)13 on crime procedure aspects, specific to information technology,
- resolution adopted by European ministers of justice, which recommend to Committee of Ministers the supporting of European Committee in Criminal Issues, so as to fight against the information criminality, by a harmonization of national legal provisions [1],
- the action plan adopted by states presidents and by governments or members of European Council, reunited at the second Summit of Strasbourg, so as to identify common solutions for implementing the standards and values of the European Council; these aspects refers to the perspective of developing and adopting new information technologies, thus being materialized by an international action under "Convention on information criminality". Not too far, on 01.28.2003, the Council of Europe issued an addendum named "Additional Protocol to Convention on information criminality, regarding the action crimes of racist or xenophobe nature, carried by means of information systems", which supplements the lacks detected after performing the Summit of Strasbourg
- recommendation R(89)9 concerning some rules that have to be applied by the member states, so as to fight against information criminality.

One of the most important Recommendations is the last issued, since in accordance to determined concepts, it aims towards treating the illegal actions related to information systems. Recommendation R(89)9 mentioned two classes of illicit facts, meaning: the minimal list and the optional list.

The two lists include behaviors that generate the phenomenon of information criminality.

Therefore, the fraud in informatics, the counterfeit in informatics, the prejudices brought to data or programs for the computer, the information sabotage, the unauthorized access, the unauthorized reproduction of protected computer programs, the unauthorized reproduction of protected topography represent the minimal list, and the alteration of data and computer programs, the information espionage, the unauthorized utilization of a protected computer program are suggesting the optional list [2].

In this way, Recommendation R(89)9 plays a definitive part in erosion of information criminality phenomenon, and determines suggestive goal towards the state entities on manifesting the adaptability in filling these lists with the following: creation and spreading of information viruses, thus influencing the good operability of it and creating prejudices to information companies.

Due to the international jurisprudence as concerns the information criminality, the European Union has approached on issuing some directives related to the field of information technology:

- Directive 2006/58/EC of 27 June 2006 amending Council Directive 2002/38/EC, as regards the period of application of the value added tax arrangements, applicable to radio and television broadcasting services and certain electronically supplied services;
- Directive 2006/24/EC of the European Parliament and of the Council 15 March 2006 on the retention of data generated or processed in connection with the provision of publicly available electronic communications services or of public communications networks and amending Directive 2002/58/EC (Data retention directive);
- Directive 2004/48/EC of the European Parliament and of the Council of 29 April 2004 on the enforcement of intellectual property rights;
- Directive 2003/98/EC of 17 November 2003 of regards the reutilization of public field information;
- Directive 2002/38/EC of 7 May 2002 of modifying and temporary changing of Directive 77/388/CEE, as regards regime of application of the value added tax arrangements, applicable to radio and television broadcasting services and certain electronically supplied services;
- Council Regulation (EC) no. 792/2002 of 7 May 2002, of temporary modification of Regulation (EEC) no. 218/92, as concerns the administrative consideration within indirect taxation (VAT) on additional measures of electronic commerce;
- Directive 2001/29/EC of 22 May 2001 on the harmonization of some aspects specific the copyrights and the connected rights to information companies;
- Directive 2000/46/EC of 18 September 2000, as regards the initiation, the exertion and prudential supervision of activities related to electronic money institutions;
- Directive 2000/31/EC of 8 June 2000, as regards judicial points of the information companies services, especially the electronic commerce on the internal market (directive regarding the electronic commerce);
- Directive 1999/93/EC of 13 December 1999, regarding a community's frame specific to electronic signatures;
- Directive 1997/7/CE of 20 May 1997 concerning the protection of users, when remote deeds conclusion is applied;
- Directive 1995/46/CE of the European Parliament and of the Council of 24 October 1995 on the protection of individuals with regard to the processing of personal data and on the free movement of such data.

The legislative activity specific to fighting against the criminality phenomenon has become a significant concerning of the international organizations, so that year 2002 brought a series of directives that refer to electronic communications: Directive 2002/58/EC, Directive 2002/19/EC, Directive 2002/20/EC, Directive 2002/21/EC, Directive 2002/22/EC in Europe. One might emphasize the fact that USA legislation has approached this crime frame since 1976 by the "Copyright Act" 17 of USC Sec. 101 et seq. (1976), "Privacy Protection Act "42 U.S.C. Sec. 2000aa (1980), "Electronic Communications Privacy Act" (1985), where year 2003 improved the legislative frame by "CAN-SPAM Act" of 2003.

Thus, the adhesion towards a unilateral consensus of Romania has also been included to the states group that took into account directives, so as to adopt them at national level. In this way, the legislative directives included some clear series of actions on defining the crime's activities, related to information networks: the electronic commerce, the copyright, the electronic signature, the electronic payments, the online advertisement, the private life (protection of private data), information criminality, pornography on the Internet or electronic communications.

The legislation concerning the information criminality can be found by means of:

- Disposals as regards the prevention and fighting against the information criminality (Title III of Law 161 of 19/04/2003, related to some measures of ensuring the transparency on exerting public stateliness, the public functions and business environment, preventing and sanctioning the corruption – Published in an official newspaper named Monitorul Oficial, Part I, no. 279 of 21/04/2003);
- Law no. 64 of 24/03/2004 in order to approve the Convention of European Council, as regards the information criminality, adopted at Budapest on 23 of November 2001;
- Law 298/2008 as concerns the retention of data generated or processed by the providers of electronic communications services, specific to public or public communication networks, as well as modifying the Law no. 506/2004 as concerns the processing of personal data and the protection of private life within electronic communications sector.

## 3 CRIMES CONCERNING THE NATIONAL INFORMATION CRIMINALITY

In Romania, one of the most important judicial regulations is represented by Law no. 161/2003, as concerns some measures of ensuring the transparency and exerting the public stateliness, the public functions and the business environment, the prevention and punishing the corruption [3]. This normative act introduced seven crimes that correspond to the international laws, and "Prevention and fighting against information criminality" might be seen in Title III of the law.

According to the legislative frame on Law no. 161/2003, one might notice three types of crimes:

a) Crimes against the confidentiality and integrity of data and information systems:
- the crime of unauthorized access to an information system,
- the crime of illegal interception of information data transmission,
- the crime of modifying the integrity of information data,
- The crime of performing illegal operations by means of illegal operations or information programs.

b) Information crimes:
- the crime of information counterfeit,
- The crime of information fraud.

c) The infantile pornography by means of information systems.

In order to identify the elements of crimes specific to information criminality, as regards the crime point of view, one might study and perform approached over the crimes concerning the information counterfeiting:

- *the legal content*, represented by art. 48 of Law 161/2003 and art. 445 of Penal code, meaning: "the fact of introducing, modifying or deleting with no rights the information data or the fact of restriction, without authorization, the access to this data, if the actions will result in achieving data not corresponding to the truth, in order to be used on producing a judicial outcome";

- *the pre-existing conditions.*

### 3.1 The crime objective

a) *the special judicial objective* – consists of social relationships regarding the public trust in security and viability of information systems, the authenticity of information data and the security of data storing and processing, or automatic transaction of data on private or official interest;

b) *the material objective* – information data under the form of alphanumerical characters, which are displayed on a monitor or printed, which might be changed.

### 3.2 The crime subjects

a) *the active subject* – might be represented by any person initiated in computer science, which by means of working activities, might have the access to data and information systems. The participation to this might be represented by co-authors, instigation or complicity;

b) *the passive subject* - might be that judicial person, which was prejudiced of own interests, by counterfeiting the information data.

### 3.3 The constitutive content

a) *the objective part*

- the material element – refers to that performed by an alternative action of introduction, modification or deleting the information data or restricting the access to this data [4]. The activity by which the material crime element is carried out will generate negative effects, detrimental to data and especially, over their operational ability, attesting situations that create errors on documents or totally counterfeiting of these. Counterfeiting the data can be done by:

- insertion, modification or deleting of data within data basis existing at the central level (electronic evidence of a bank, for instance), by direct action of the accomplisher to the keyboard or by means of an external device;

- the alteration of documents stored electronically, by means of introducing errors or by deleting or modifying them.

In this way, the information counterfeiting might take one of the aspects: feigning the electronic mail, feigning of hyper-connections, feigning of web, SPOOFING feigning [5], the attack by dissimulation.

- *the immediate result* - refers to the significant aim, of achieving data not corresponding to the truth and in consequence, creating of a dangerous status, by trust given to information data and in generally, to automatic processing of data;

- *the causality connection* – the activity of accomplisher and the results will create a prejudice that has to be proven.

b) *the subjective part*. The information crime has been accomplished by immediate intention, an intention determined by its aim. The goal is that of using inadequate data, in the view of producing a judicial consequence. The data are susceptible on producing judicial results, if is able to arise, modify or even delete the judicial reports, thus creating rights and obligations [6].

### 3.4 Forms. Means. Sanctions

a) *forms* – the actions preparing the crimes are not sanctioned, but the attempt is punished according to art. 50 of La1 161/2003. The crime is considered accomplished, when the accomplisher has done the modification of data, by insertion, changing or by restricting the information data;

b) *means* – the main possibilities consists of: introducing erroneous information data, deleting of data, modifying of data or restricting the access to data;

c) *sanctions* – the punishment, in according to the national legislation on arresting during a time between two to seven years.

### 3.5 Procedural aspects

The penal action is started ex-officio.

## 4 CONCLUSIONS

Taking into consideration the analysis of disposals on penal character, one might emphasize that for the time being, there is a coherent frame by which an effective protection of social values in this field exists, by accomplishing the final aim on preventing the accomplishment of crimes, related to information technology. Although, the continuous progress of the information technology has to be always taken into account, so that any products might support improvements, products either specific to the human activity or those related to a legislative frame within penal protection field on integrity and security of data and information systems.

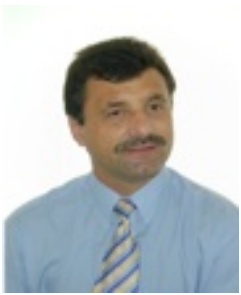

Constantin Filote - Associate professor, PhD, Faculty of Electrical Engineering and Computer Science, Department of Computers and Automation, 'Stefan cel Mare' University of Suceava. Author of 6 specialty books within information field, 6 scientific papers published within journals of international conferences and included in international data basis, 10 scientific papers published in specialty journals and ISI Thomson quoted, included in international data basis, member of 5 professional associations, director of 5 specialty projects and member of 14 project teams, coordinator of 5 international conferences.

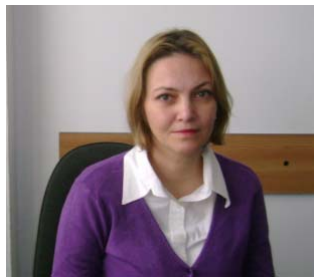

Nemţoi Gabriela - Lecturer, PhD candidate, Faculty or Economics and Public Administration, Department of Law and Judicial Sciences, 'Stefan cel Mare' University of Suceava. Author of 30 papers, participant to 5 international conferences, with papers included to journals, 4 scientific papers published in data basis and EBSCO included, PhD Candidate at State Academy of Republic of Moldova. The PhD thesis title: "The part of people's sovereignty over state organization of capacities".